\documentclass[aps]{revtex4}
\usepackage{color}
\usepackage{graphicx}
\usepackage{eurosym}
\usepackage{graphicx}
\usepackage{amsmath,amssymb}
\usepackage{color}
\usepackage{amsmath}
\usepackage{amsfonts}
\usepackage{amssymb}

\def\bc{\begin{center}}
\def\ec{\end{center}}

\def\beq{\begin{eqnarray}}
\def\eeq{\end{eqnarray}}

\def\bc{\begin{center}}
\def\ec{\end{center}}

\def\beq{\begin{eqnarray}}
\def\eeq{\end{eqnarray}}

\input{tcilatex}

\begin{document}

\title{Achieving the required mobility in the solar system through Direct
Fusion Drive}
\author{Giancarlo Genta$^{1}$ and Roman Ya. Kezerashvili$^{2,3,4}$,}
\affiliation{\mbox{$^{1}$Department of Mechanical and Aerospace Engineering, Politecnico di
Torino, Turin, Italy}\\
\mbox{$^{2}$Physics Department, New York City College
of Technology, The City University of New York,} \\
Brooklyn, NY, USA \\
\mbox{$^{3}$The Graduate School and University Center, The
City University of New York,} \\
New York, NY, USA \\
\mbox{$^{4}$Samara National Research University, Samara, Russian Federation}\\
}
\date{\today }

\begin{abstract}
To develop a spacefaring civilization, humankind must develop technologies
which enable safe, affordable and repeatable mobility through the solar
system. One such technology is nuclear fusion propulsion which is at present
under study mostly as a breakthrough toward the first interstellar probes.
The aim of the present paper is to show that fusion drive is even more
important in human planetary exploration and constitutes the natural
solution to the problem of exploring and colonizing the solar system.
\end{abstract}

\maketitle
\section*{Nomenclature}

$I_{s}$\qquad specific impulse

$m$\qquad mass

$m_{i}$\qquad initial mass

$m_{l}$\qquad mass of payload

$m_{p}$\qquad mass of propellant

$m_{s}$\qquad structural mass

$m_{t}$\qquad mass of the thruster

$m_{tank}$\qquad mass of tanks

$t$\qquad time

$t_{d}$\qquad departure time

$v_{e}$\qquad ejection velocity

$F$\qquad thrust

$J$\qquad cost function

$P$\qquad power of the jet

$\alpha $\qquad specific mass of the generator

$\gamma $\qquad optimization parameter

$\Delta V$\qquad velocity increment

DFD\qquad Direct Fusion Drive

IMLEO\qquad Initial Mass in Low Earth Orbit

LEO\qquad Low Earth Orbit

LMO\qquad Low Mars Orbit

NEP\qquad Nuclear Electric Propulsion

NTP\qquad Nuclear Thermal Propulsion

SEP\qquad Solar Electric Propulsion

SOI\qquad Sphere of Influence

VEV\qquad Variable Ejection Velocity

\section{Introduction}

To develop a spacefaring civilization, humankind must develop technologies
which enable safe, affordable and repeatable mobility through the solar
system. Half a century ago (the last year we were celebrating the $50^{th}$
 anniversary of the first human landing on the Moon) we have proven that the
technology then (and today, since little has changed in this field in the
last 50 years) available was barely sufficient to reach the closest
celestial body, -- the Moon, for a number of flag and footprint missions.
Although never yet attempted, it is possible to assess that the same
technology can allow to perform some preliminary human missions to Mars \cite%
{1,2}.

Although it is well known that to really explore and colonize the closest
celestial bodies a wide range of technologies need to be developed \cite{3}
-- technology to exploit in-situ resources, to protect the astronauts from
radiation, to build manufacts on the destination planet, etc. -- new
technologies directly related to propulsion are required. In particular,
it is essential to use nuclear energy instead of chemical energy to propel
spacecraft.

Both alternatives of Nuclear Thermal and of Nuclear Electric Propulsion (NTP
and NEP) based on nuclear fission reactions have been studied in detail, and the former was already bench
tested with very satisfactory results. NTP and NEP can allow to improve our
chances of performing human missions to Mars and beyond by reducing the
travel time (and thus the exposure of the crew to cosmic radiation) while at
the same time reducing the Initial Mass in Low Earth Orbit (IMLEO) and thus
making interplanetary missions more affordable. An interesting comparison
between the NTP and chemical approach to a human Mars mission is reported in
the NASA Design Reference Architecture 5 (DRA5) \cite{3,4}.

Also NEP allows a notable improvement with respect to chemical propulsion,
and the choice between the two mentioned nuclear approaches depend mainly on
political decisions about which technology to develop to a sufficient
Technology Readiness Level. Both the mentioned nuclear approaches are based
on fission nuclear reactions \cite{5}.

Recent advances in light weight structures and thin film solar cells make it
possible to think of using Solar Electric Propulsion (SEP) also for human
planetary missions and in particular for the first human missions to Mars.
This is a sort of 'bridge' solution to improve the performance of
interplanetary spacecraft above those of chemical propulsion, while waiting
that the technology for NTP or NEP becomes available. By comparing the
performance of SEP with that of chemical propulsion and NTP, the advantages
in terms of IMLEO are clear, while with respect to NEP they depend only on
the specific mass of the generator $\alpha $, which in the short term is
more favorable for solar arrays than for nuclear generators. In a longer
term, the latter will be much better, but developing SEP means developing
high power electric thrusters for human missions so that they will be ready
when lightweight nuclear generator will become available.

At any rate there is no doubt that to become a real spacefaring civilization
we must develop rocket engines based on nuclear fusion \cite{6,7}. The idea
to use fusion power for spacecraft propulsion has a long history \cite%
{Angelo}. For the fusion propulsion there are two alternatives:similar to NTP and fusion NEP. In the last 20 years
many studies have been devoted to the development of fusion nuclear power in
general -- mostly for general power generation -- and specifically of fusion
nuclear rockets. Fusion NEP requires the development of lightweight fusion
reactors, which is something that today appears to be a difficult
achievement. Moreover, also here the point is again just the specific mass
of the generator $\alpha $, and many years will pass before fusion generator
will have a better value of $\alpha $ than fission generators \cite{8} --
apart from the fact that today no fusion generator, even with a very high $%
\alpha $, is available. In fusion NEP, the lower is the value of $\alpha $, the
higher is the optimum value of the specific impulse, so even when a
lightweight generator will be available, much work will be required also for
improving the electric thruster.

The revolutionary Direct Fusion Drive (DFD) is a nuclear fusion engine and its concept is based on the Princeton field-reversed configuration reactor, which has the ability to produce thrust from fusion without going through an intermediary electricity-generating step \cite{17}. The engine development is related to the ongoing fusion research at Princeton Plasma Physics Laboratory. The DFD uses a novel magnetic confinement and heating system, fueled with a mixture of isotopes of helium and hydrogen nuclei, to produce a high specific power, variable thrust and specific impulse, and a low-radiation spacecraft propulsion system. The simplest type of fusion drive is using small uncontrolled thermonuclear
explosions to push forward the spacecraft, as was planned in the Orion
Project \cite{5}, but even if a continuous, controlled reaction is used, DFD
seems to be much easier to realize and D $-$ $^{3}$He direct fusion
thrusters seem to be the thrusters which will allow to colonize, in the
medium term, the solar system.

While most of the studies related to DFD deal with missions to the outer
solar system or the near interstellar space, the aim of the present paper is
studying in some detail fast human missions to Mars and to the Asteroid
Belt. The result is that nuclear fusion propulsion is the enabling
technology to start the colonization of the solar system and the creation of
a solar system economy.

The paper is organized in the following way: In Section II we describe the thruster and its main characteristics.
Section III is devoted to considerations of three cases for the Earth -- Mars mission: i. the ideal variable ejection velocity (VEV) operations; ii. limited VEV operations;
iii. slow cargo spacecraft mission. The mission to 16 Phyche asteroid is considered in Section IV, and finally conclusions are given in Section V.

\section{The thruster}

The solar system exploration and beyond, from robotic space voyages to
manned interplanetary missions, requires high-thrust, high-exhaust velocity
engines. Our experience suggest that any engine engineered should be based
on the well-understood physics of today. The emphasis on known physics and
affordability limits the scope still further to nuclear processes: fission
and fusion. However, both fission- and fusion-based propulsion schemes, are
well understood and realizable at power levels, while at the same time the
mass of fuel, propellant, structure, and shielding severely limit their
space flight capabilities. A survey of rocket-engine performance for solar
system missions beyond the Moon-Earth system has compared chemical and
nuclear fission and fusion power sources \cite{9}. One conclusion reached is
that chemical rockets have reached their practical limits, epitomized by
long-duration, low-payload-mass missions. A corollary is that nuclear power
will be needed for more ambitious missions \cite{10}.

A nuclear fission reactor is used in a rocket design to create Nuclear
Thermal Propulsion. In an NTP rocket the type of nuclear reactor, is ranging
from a relatively simple solid core reactor up to a much more complicated
but more efficient gas core reactor. The fission reactions are used to heat
liquid hydrogen which flows around the fission region inside a reactor and
absorbs energy from the fission products. High temperature heating turns
liquid hydrogen into ionized hydrogen gas, which is then exhausted through a
rocket nozzle to generate thrust. As an alternative, the propellant can be
heated directly by the fission fragments, like in the thruster proposed by
Carlo Rubbia \cite{101}. The specific impulse produced is proportional to
the square root of the temperature to which the working fluid is heated. The
hydrogen propellant typically delivers specific impulses on the order of 850
to 1000 seconds, which is about twice that of liquid hydrogen-oxygen
chemical rocket. A second possible method, which relays on nuclear fission
to generate propulsion, known as Nuclear Electric Propulsion, involves the
same basic fission nuclear reactors. Heat extracted from a fission chain
reaction is converted into electrical energy which then powers an electrical
engine. One should mention that electric specific impulse thrusters
typically use much less propellant than chemical rockets because they have a
higher. Therefore, in both cases, the rocket relies on nuclear fission to
generate propulsion. The nuclear fission propulsion is limited by thermal
inefficiencies and that fusion could provide more and better mission options
because of its higher power conversion efficiency and higher energy-content
fuel \cite{9}.

Fusion reactions produce much more energy than fission processes. Usually
one of the components of the fusion reaction is protium (hydrogen atom
without any neutron), deuterium (hydrogen atom with proton and neutron), or
tritium (hydrogen atom with proton and two neutrons). The other component
which involves into the fusion process of light nuclei can be deuterium,
isotopes of helium, $^{3}$He or $^{4}$He, and isotopes of lithium, $^{6}$Li
and $^{7}$Li. In fusion reactors use the energy released by the fusion of
light atomic nuclei. Let us focus of the fusion processes in the
deuterium--deuterium (D--D), deuterium--tritium (D--T) and deuterium--$^{3}$%
He (D--$^{3}$He) plasma.

The D--D plasma admits the following primary reactions:
\begin{eqnarray}
\text{D}+\text{D} &=&\text{ }^{3}\text{He (0.82 Mev)}+n\text{ (2.45 Mev)}%
+3.25\text{ MeV,}  \label{DD3He} \\
\text{D}+\text{D} &=&\text{T (1.01 Mev)}+p\text{ (3.03 Mev)}+4.04\text{ MeV.}
\label{DDT}
\end{eqnarray}
In Eqs.~(\ref{DD3He}) and (\ref{DDT}) the values in parenthesis are the
energy of that particular fusion product. The D--T plasma admits both
deuterium--deuterium reactions (\ref{DD3He}) $-$ (\ref{DDT}) and
deuterium--tritium processes. The primary energy reactions in the D--T
plasma in addition to (\ref{DD3He}) and (\ref{DDT}) are the following:

\begin{eqnarray}
\text{D}+\text{T} &=&\text{ }^{4}\text{He (3.52 Mev)}+n\text{ (14.06 Mev)}%
+17.6\text{ MeV,}  \label{DT} \\
\text{T}+\text{T} &=&\text{ }^{4}\text{He (3.52 Mev)}+2n\text{ }+11.3\text{
MeV.}  \label{TT}
\end{eqnarray}%
There are two reactions of importance in the D--T plasma: the D + D and D +
T. The first one is the reaction between two nuclei of deuterium. The second
occurs in the interaction of deuterium with tritium. In both cases fast
2--14 MeV neutrons are emitted, which should be shielded to protect the
spacecraft and crew. Energetically the most preferable process is D + T
fusion, which releases 17.6 MeV. The primary reaction is D + D has two
almost equally exothermic channel (\ref{DD3He}) and (\ref{DDT}). They lead
to the development of the plasma active isotopes and that initiates the
emergence of important secondary catalytic processes (\ref{DT}) and (\ref{TT}%
).

However, the cross section for (\ref{DD3He}) and (\ref{DDT}) reactions at
low energies is considerably smaller than the cross-sections for the D--T
process (\ref{DT}), and its ignition requires much higher temperatures of
plasma. The ideal temperature threshold is $\sim $40 keV (about 5$\times $10$%
^{8}$ K).

The fusion reaction of nuclei of tritium and deuterium is the most promising
for the implementation of controlled thermonuclear fusion, since its cross
section even at low energies is sufficiently large \cite{11}. However, the
problem of the contamination due to the neutron emission in the primary
processes still exists. Although the D--D fuel burning is also accompanied
by a neutron flux, they are weakened in comparison with the D--T process (%
\ref{DT}).

Now let us consider the reaction that are admitted in D--$^{3}$He plasma.
The primary processes which occur in D--$^{3}$He plasma are the aneutronic
fusion
\begin{equation}
\text{D}+\text{ }^{3}\text{He}=\text{ }^{4}\text{He (3.52 Mev)}+p\text{
(14.7 Mev)}+18.34\text{ MeV}  \label{D3He}
\end{equation}%
and reactions (\ref{DD3He}) and (\ref{DDT}) that involve the D--D fusion.
One should mention that reaction (\ref{D3He}) is the secondary process in
the D--T plasma. Due to the D--D fusion, the D--$^{3}$He plasma also
includes undesired neutron channel (\ref{DD3He}). The two branches of the
D--D fusion reaction produce $^{3}$He and tritium. The only neutrons
produced are medium-energy neutrons (2.45 MeV). If the produced $^{3}$He
(Helium-3) is not removed, it will react with deuterium producing charged
particles protons and $^{4}$He with no additional neutrons. On the other
hand, if the produced tritium is not removed, it will react with deuterium
producing 14.1 MeV neutrons. However for the D --$^{3}$He plasma with equal
deuterium and Helium-3 densities, the fraction of the fusion energy carried
by neutrons from the D--D reaction is 1/3 \cite{10, 12}. The undesired
tritium produced via reaction (2) will increase the neutron flux due to
reactions (\ref{DT}) and (\ref{TT}), which are secondary for D--$^{3}$He
plasma. In Refs. \cite{13, 14} a method of tritium removal from the plasma
before it can fuse has been proposed. Removing tritium produced by D--D
fusion and recycling part of it after it decays to Helium-3 isotope significantly
reduces the fraction of fusion energy carried by neutrons in a D--D system
\cite{14}. The latter results in significant lifetime enhancement of
structural materials. In summary, the main reactions in D--$^{3}$He fusion
produce far fewer neutrons than D--D fusion. Consequently, a lower mass of
shielding materials is required, which will reduce the total mass of the
structure. Calculations show that for obtaining useful energy the
temperature of ions in a plasma for the D + D reaction should be about 10$%
^{9}$ K and for the D + T reaction about 10$^{8}$ K \cite{15}.

The experimental study of the D--$^{3}$He plasma led to the proposal of a
new kind the fusion-based thruster - the Direct Fusion Drive (DFD), which
has field-reversed configuration (FRC) reactor for an original
plasma-formation. The FRC employs a linear solenoidal magnetic-coil array
for plasma confinement and operates at higher plasma pressures. One should
note that several FRCs \cite{Jones,Cohen2007,Petrov,Guo} have achieved
stable plasmas. The DFD employs the radiofrequency technique called rotating
magnetic field (RMF) to form and heat plasma. An important figure-of-merit
for fusion reactors is $\beta $, the ratio of the plasma pressure to the
magnetic energy density. The innovative radiofrequency RMF method, which
heats particles and allows the size of the FRC to be relatively small was
suggested in Ref. \cite{Cohen2000}. In Refs. \cite{10, 12, 16, 17} was
considered a compact, anuetronic fusion engine, which enables more
challenging exploration missions in the solar system and beyond. The DFD
concept is result of the Princeton Field-Reversed Configuration Reactors
which employ heating method invented by S. Cohen. The Scrape-off Layer (SOL)
of the DFD is quite different than that of any other fusion device. The
energy is deposited in the SOL directly from the D--$^{3}$He fusion products
via a non-local process and is predominantly transmitted to the electrons
via fast-ion drag. The random thermal energy in the SOL electrons is
transferred to the cool SOL ions through a double layer at the nozzle and
via expansion downstream, thus being converted into directed flow of a
propellant fluid. The heat transport into SOL is described by Fick's law
\cite{Fick}, by the local flux-surface-normal gradient in pressure. In Refs.
\cite{17,ThomasOhio} is used a fluid model for the SOL between the gas box
with the propellant and the nozzle and dependencies of the thrust and
specific impulse on gas input flow for powers of 0.25 to 7 MW transferred to
the SOL are studied. The calculations are performed using UEDGE \cite{UEDGE}
fluid-code for simulations. Results of these simulations for the propellant gas
input 0.08 g/s yields to the data given in Table \ref{Tab0}. The nuclear
fuel for such a thruster is D--$^{3}$He and the propellant fluid is atomic
or molecular deuterium, which is heated by the fusion products and then
expanded into a magnetic nozzle, generating an exhaust velocity and thrust.
Adding propellant to this flow results in a variable thrust, variable
specific impulse exhaust through a magnetic nozzle. The thrust of the DFD
depends on the input gas flow and varies from about 4 N for the power 0.25
MW to 60 N, when the power is 7 MW. The results of simulations show that
when the gas input flow increases from 0.08 to 0.7 g/s the thrust increases
from 4 N to 60 N. For the specific impulse the preferable gas feed for the
power 0.25 MW to 7 MW is 0.08 -- 0.3 g/s \cite{17,ThomasOhio}. Approximately
35\% of the fusion power goes to the thrust, 30\% to electric power, station
keeping and communication, 25\% lost to heat, and 10\% is recirculated for
the radio frequency heating. The current estimated DFD specific powers are
between 0.3 and 1.5 kW/kg \cite{17}. One could considered as a conservative
estimate the power to thrust efficiency, about 0.3 -- 0.5.

\begin{table}[t]
\caption{DFD propulsion parameters based on UEDGE Model spacecraft based on
the DFD studies in Refs. \protect\cite{17,ThomasOhio,Thomas2019}}
\label{Tab0}
\begin{center}
\begin{tabular}{cccc}
\hline\hline
Total Fusion Power, MW & $1$ & $1.5$ & $2$ \\
Specific Impulse, $I_{s}$, s & $10,000$ & $12,000$ & $12,500$ \\
Thrust, $T,$\ N & $8$ & $10$ & $11$ \\
Fusion Efficiency & \multicolumn{3}{c}{$0.17-0.46$} \\
Specific Power, kW/kg & \multicolumn{3}{c}{$0.75-1.25$} \\ \hline\hline
\end{tabular}%
\end{center}
\end{table}

A full-sized D--$^{3}$He fusion reactor is perfectly suited to use as a
rocket engine for two reasons:

\qquad i$.$ the configuration results in a radical reduction of neutron
production compared to other D--$^{3}$He approaches;

\qquad ii. the reactor features an axial flow of cool plasma to absorb the
energy of the fusion products. In other words, the cool plasma flows around
the fusion region, absorbs energy from the fusion products, and then is
accelerated by a magnetic nozzle.



\begin{table}[b]
\caption{Main characteristics of the Discovery II and a spacecraft based on
the DFD thruster studied in Ref. \protect\cite{17}}
\label{Tab1}
\begin{center}
\begin{tabular}{ccc}
\hline\hline
Spacecraft & Discovery II & DFD \cite{17} \\ \hline
Specific impulse $I_{s},$ s & 47,205 & 23,000 \\
Specific mass of the propulsion system $\alpha ,$ kg/W & 0.000116 & 0.0018
\\ \hline\hline
\end{tabular}%
\end{center}
\end{table}

The ratio of the total mass of fuel and propellant to the mass of $^{3}$He
is about 670 and to get a specific power 0.18 kW/kg the small amount of $%
^{3} $He is needed -- about 0.53 kg. The most Helium-3 used in industry
today is produced from the radioactive decay of tritium. Tritium is a
critical component of nuclear weapons and it was produced and stockpiled
primarily for this. At present, Helium-3 is only produced as a byproduct of
the manufacture and purification of tritium for use in nuclear weapons. The
main source of Helium-3 in the United States is the federal government's
nuclear weapons program \cite{18}. There are extraterrestrial sources of $%
^{3}$He. Materials on the Moon's surface which contains Helium-3 at
concentrations between 1.4 and 50 parts-per-billion in sunlit and shadowed
areas \cite{Wittenberg,19, 20}. There is may be Helium-3 on Mars also \cite%
{Wittenberg2}. The analysis of fusion fuel resource base of our solar system
is given in Ref. \cite{Robert}.

A spacecraft driven by a fusion thruster was studied at the turn of the
century by NASA: its goal was to perform a human mission to Jupiter or
Saturn as described in the Movie 2001, a Space Odyssey. The spacecraft was
aptly named Discovery II \cite{21}. The main characteristics of the thruster
(in the version for the Saturn mission) which were obtained in that study
were: specific impulse $I_{s}$ = 47,205 s and specific mass of the
propulsion system $\alpha =$ 0.00016 kg/W and are listed in Table \ref{Tab1}. These values are very favorable
indeed. DFD, with a specific impulse in the range of 10,000 to 20,000
seconds and a specific power about 1 kW/kg, is suitable for almost any
interplanetary mission. In NASA solicitation for rapid, deep space
propulsion, four candidate missions were identified: Mars, Jupiter, Pluto,
and 125 AU for an interstellar precursor mission. In Ref. \cite{Thomas2019}
has sized a DFD engine for each candidate mission. The main characteristics
of the spacecraft are reported in Table \ref{Tab0}. A more recent study for
a DFD driven spacecraft is reported in Ref. \cite{17}. Although showing a
small spacecraft aimed at the focal line of the gravitational lens of the
Sun and an interstellar probe aimed at Alpha Centauri, the basic values
there reported for a 2 MW fusion rocket can be considered as a conservative
estimate for a larger unit aimed to power an interplanetary piloted
spacecraft.

\section{Earth -- Mars Mission}

\subsection{Ideal Variable Ejection Velocity operations}

In our consideration of the Earth -- Mars mission we use the parameters for the DFD thruster given in Table \ref{Tab1}. As it was shown by the authors in a previous paper \cite{Genta}, a thruster
with such a high specific impulse and low specific mass must operate in a
continuous thrust mode. A first study of an Earth-Mars transfer was
performed assuming that it can operate in an optimal (unlimited) Variable
Ejection Velocity (VEV) conditions. The study was performed using the IRMA
7.1 computer code \cite{22} with the following data: launch opportunity:
2037; specific mass $\alpha =1.25$ kg/kW; overall efficiency $\eta =0.56$;
tankage factor $k_{tank}=0.10$; height of circular starting LEO: 600 km;
height of circular arrival LMO: 300 km.

\begin{table}[b]
\caption{Timing and mass breakdown of the missions studied in the present
paper.}
\begin{center}
\begin{tabular}{lcccc}
\hline\hline
Destination & \multicolumn{3}{|c|}{Mars} & Psyche \\
Type & Unlimited & Limited fast & Limited cargo &  \\ \hline
$t_{d}$ (days) & 66.6 & 71.9 & 189.9 & 120 \\
$t_{t}$ (days) & 120 & 123 & 350 & 250 \\
$t_{1}$ (days) & 8.4 & 20.6 & 94.5 & 27.6 \\
$t_{2}$ (days) & 105.8 & 96.0 & 219.2 & 222.3 \\
$t_{3}$ (days) & 5.8 & 6.4 & 36.3 & 0.1 \\
$(m_{l}+m_{s})/m_{i}$ & 0.254 & 0.258 & 0.715 & 0.241 \\
$m_{p}/m_{i}$ & 0.525 & 0.319 & 0.178 & 0.416 \\
$m_{t}/m_{i}$ & 0.169 & 0.391 & 0.089 & 0.169 \\
$m_{tank}/m_{i}$ & 0.052 & 0.0319 & 0.018 & 0.042 \\
$P_{jet}/m_{i}$ (W/kg) & 75.62 & 175.35 & 40.09 & 134.60 \\ \hline\hline
\end{tabular}%
\end{center}
\label{Tab21}
\end{table}

The optimal trajectory for a 120 days Earth-Mars journey starts 66.6 days
before opposition, spends 8.4 days spiraling about Earth, 105.8 days in
interplanetary space and finally 8.4 days spiraling about Mars to reach the
final LMO. The mass breakdown and the jet power are reported in Table \ref%
{Tab21}, second column. Notice that the ratio between the installed power
and the vehicle mass is of the order of magnitude of that of a modern small
car, showing that traveling fast in the solar system does not require
enormous amounts of power!

\begin{figure}[h]
\noindent
\begin{centering}
\includegraphics[width=8.0cm]{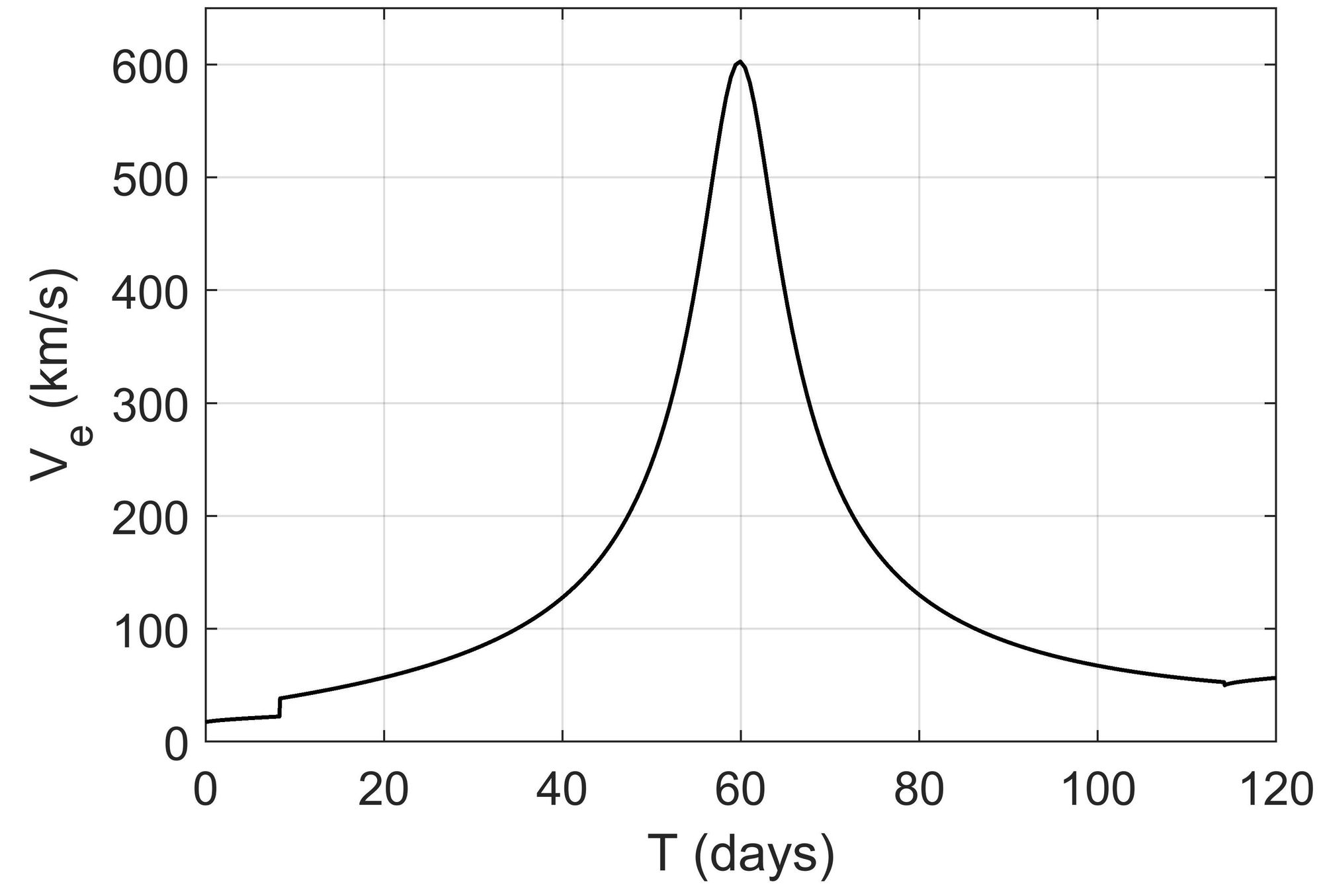}
\par\end{centering}
\caption{(Color online) Time history of the Specific
impulse during a 120 days Earth -- Mars transfer.}
\label{Fig1}
\end{figure}

The optimum specific impulse is shown in Fig. \ref{Fig1}. The specific
impulse ranges between 1,790 s at start and 60,250 s at midcourse, which is
60 days after starting.


\subsection{Limited Variable Ejection Velocity operations}

The minimum and the maximum values of the specific impulse are certainly
beyond the possibilities of the thruster, so the computation was repeated
limiting the specific impulse between 9,900 and 12,000 s. In this case the
optimal strategy is increasing the duration of the planetocentric phases (the
specific impulse is higher than the optimal one in these phases, and keeping
their duration at the optimal value of the unlimited case would result in an
unacceptable increase of the jet power) and introducing a coast arc at the
interplanetary mid course, introducing a bang-bang regulation of the
thruster.

\begin{figure}[h]
\noindent
\begin{centering}
\includegraphics[width=11cm]{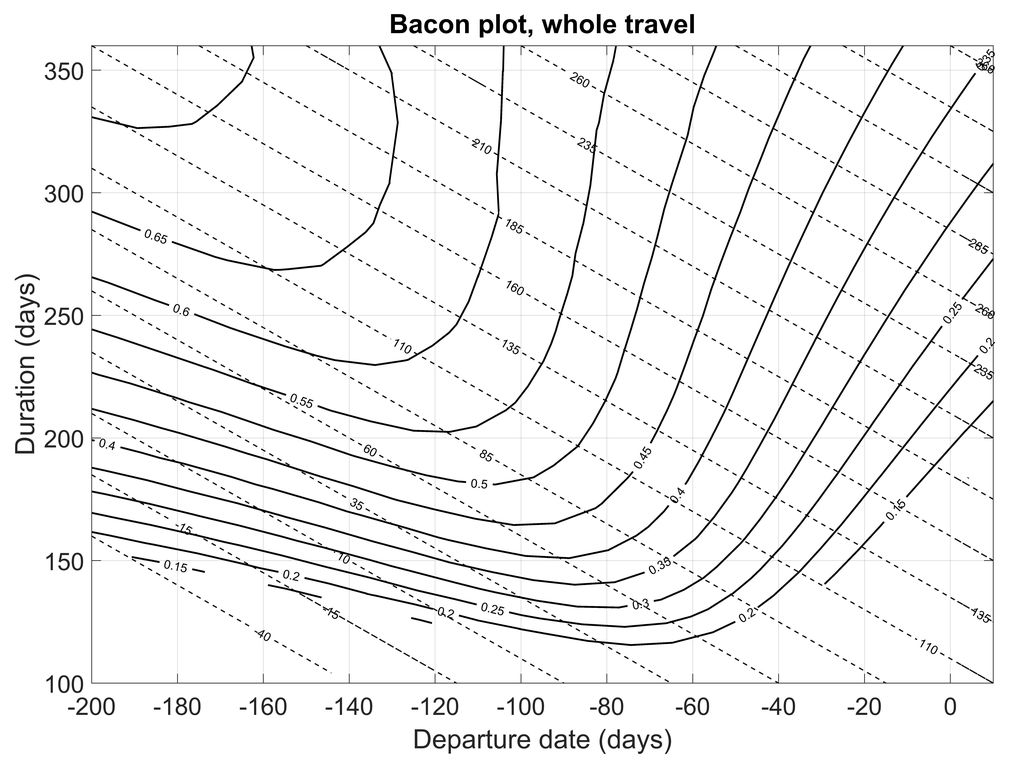}
\par
\end{centering}
\caption{(Color online) Bacon plot, i.e. the contour plot
of the surface $(m_{l}+m_{s})/m_{i}$ as a function of $t_{d}$ and $t_{tot}$,
for an Earth -- Mars transfer.}
\label{Fig2}
\end{figure}

The orbit-to-orbit bacon plot is reported in Fig. \ref{Fig2}: at equal
transfer time the payload mass fraction is slightly lower and thus to
maintain the same payload fraction a slightly longer travel time has to be
accepted.


\begin{figure}[h]
\noindent
\begin{centering}
\includegraphics[width=9.5cm]{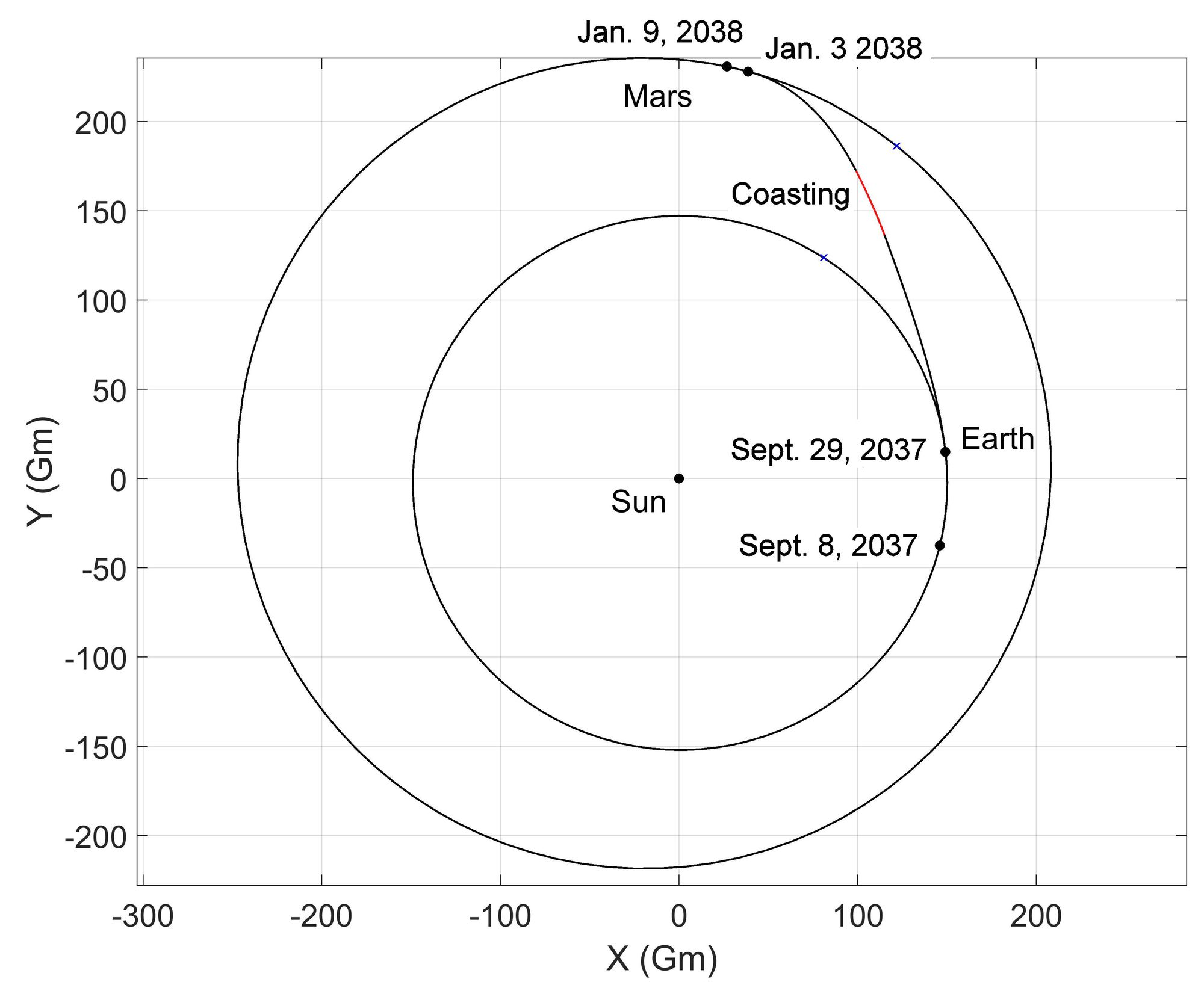}
\par
\end{centering}
\caption{(Color online) Earth -- Mars trajectory for the
fast crew spacecraft.}
\label{Fig3}
\end{figure}

A slightly longer transfer time, $t_{t}$ = 123 days, is chosen. The
trajectory starts 71.9 days before opposition. The mass breakdown and the
jet power are reported in Table \ref{Tab21}, third column. The trajectory is
shown in Fig. \ref{Fig3}, while the time histories of the acceleration, the
ejection velocity, the thrust and the power of the jet are shown in Fig. \ref%
{Fig4}

\begin{figure}[h]
\noindent
\begin{centering}
\includegraphics[width=14.0cm]{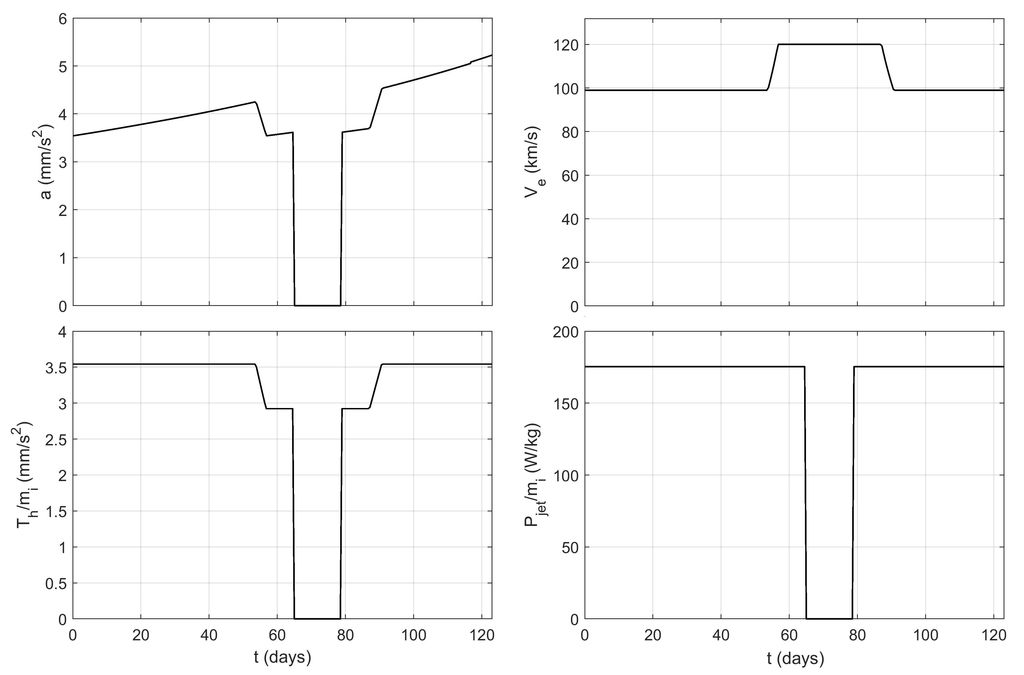}
\par
\end{centering}
\caption{(Color online) Time histories of the
acceleration, the ejection velocity, the thrust and the power of the jet
during an Earth-Mars journey $-$ fast crew spacecraft.}
\label{Fig4}
\end{figure}


The limitation of the minimum exhaust velocity reduces the propellant
consumption but causes an increase of the installed power and thus of the
mass of the thruster. The overall result is a decrease of the payload mass
at equal total journey time.

\subsection{Slow cargo spacecraft}

A slow cargo ship able to carry to Mars large payloads in an inexpensive way
can also be built with this technology. Assuming a travel time of almost one
year (namely 350 days) and starting from Earth orbit about 190 days before
the opposition, the results reported in in Table \ref{Tab21}, fourth column
are obtained. The payload and structures mass fraction is quite high, above
70\% (namely 0.715), which means that using a single superheavy launcher
able to carry 130 t in LEO, a cargo of \ about 93 t (minus the structural
mass) can be carried into LMO.

Also the power of the jet is quite small, of about 40 W/kg (referred to the
IMLEO).

\section{Mission to 16 Psyche asteroid}

Asteroid 16 Psyche, which belongs to the asteroid belt, is a metal asteroid extremely rich in nickel and iron,
but also in gold. NASA plans a mission to survey this asteroid which should
be launched in August of 2022, and arrive at the asteroid in early 2026,
following a Mars gravity assist in 2023. The asteroid has a mass of 1.7$%
\times 10^{19}$ kg and an average diameter of 226 km.

Using the DFD thruster here described a mission to the same asteroid can be
performed in a quite short time: for instance, using the launch opportunity
of 2037 (the opposition is on March 4, 2037) and starting 120 days before
the opposition, a mission lasting only 220 days can be performed. This
figure must be compared with the roughly 3.5 years of the mentioned NASA
proposal, based on chemical propulsion and gravity assist. The mass
breakdown and the jet power are reported in Table \ref{Tab21}, last column.


\begin{figure}[h]
\noindent
\begin{centering}
\includegraphics[width=9.5cm]{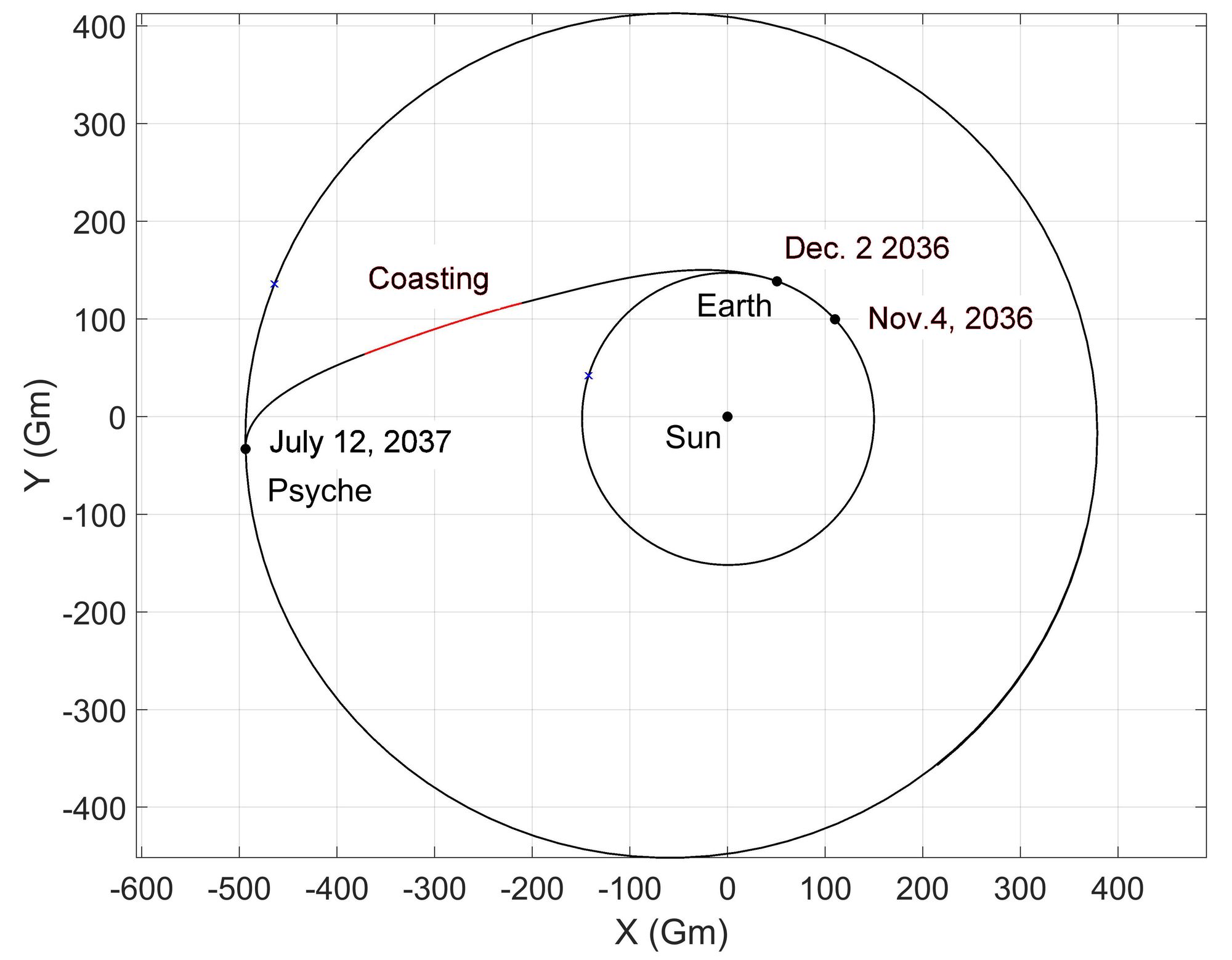}
\par
\end{centering}
\caption{(Color online) Trajectory for a 250 days
journey to the metal asteroid 16 Psyche.}
\label{Fig5}
\end{figure}

The payload and structure fraction is quite high (0.241) and the travel time
is low enough to imagine even a human mission to a metal asteroid of the
main belt like 16 Psyche $-$ since the planetocentric part of the trajectory
lasts almost one month, a human mission in which the astronauts reach the
spacecraft at the exit from the earth sphere of influence would last about
225 days, roughly like most of the human Mars missions presently planned.


\section{Conclusions}

From the study here performed it is clear that the development of a nuclear
fusion rocket engine based on the D$-^{3}$He technology will allow to travel
in the solar system with an ease never before attained, opening almost
'science fiction' possibilities to humankind.

One way travels to Mars in slightly more than 100 days become possible and
also journeys to the asteroid belt in about 250 days After the return to
Earth orbit the spacecraft can be refitted and refurbished to make another
travel in the following launch opportunity: a sort of commuting Earth-Mars
service aimed at the colonization of the red planet. A spacecraft able to
carry 30 t in 120 days or 85 t in 350 days to Mars may be launched from the
Earth surface with a single superheavy-lift launcher (slightly bigger than
the Saturn 5 or the Energia). A cargo ship able to carry to LMO the
propellant required for the return journey and much cargo is also possible.

However, the performance of such devices is still hypothetical and the value
of its specific mass here assumed is conservative also taking in mind that
this technology has very ample margins for improvements $-$ as an alternative to
chemical propulsion which has already reached the limits of this technology. If a lower value of the specific mass (a higher value of the specific
impulse) will prove to be feasible, even faster interplanetary spacecraft
could become possible.

The spacecraft described in the present paper still require much research
and development, but it is possible that they become feasible in less than
two or three decades (the launch opportunity here studied is that of 2037, -- 17 years from now), which is a fairly favorable one for Mars, while, on the
contrary is not a very good one for 16 Psyche however with such powerful
spacecraft the difference between a 'good' and a 'bad' launch opportunity is
smaller than when using chemical propulsion.

If a DFD could be made available in time for the first human Mars missions $-
$ as the launch opportunity here chosen implicitly implies $-$, the latter
would become much easier, safer and affordable than what is today thought.
These results show that the development of this technology should be given a
high priority by space agencies and public and private research centers.

Today chemical propulsion technologies are available to make a Mars mission possible and the foundation of fusion propulsion is already being built. However, it could be a fusion-powered spacecraft that ferries us to Mars in foreseeable future. We should believe that by mid-21st century, trips to Mars may become as routine as trips to the International Space Station today due to achievements in developments of fusion-powered thrusters.


\section{References}

\begin{thebibliography}{99}
\bibitem{1} R. Zubrin, The case for Mars, Touchstone, New York, 1997.

\bibitem{2} R. Zubrin, Mars Direct, Penguin Books, London, 2013.

\bibitem{3} G. Genta, Next stop Mars - The Why, How, and When of Human
Missions, Springer-Praxis, New York, 2017.

\bibitem{4} B.G. Drake ed., Mars Architecture Steering Group, Human
Exploration of Mars, Design Reference Architecture 5.0 (and addendums), NASA
Johnson Space Center, 2009.

\bibitem{5} F. Chang D\'{\i}az, E. Seedhouse, To Mars and Beyond, Fast!, How
Plasma Propulsion Will Revolutionize Space Exploration, Springer-Praxis, New
York, 2017.

\bibitem{6} G. Dyson, Project Orion, Penguin Books, London, 2002.

\bibitem{7} P.A. Czyszn, C. Bruno, Future Spacecraft Propulsion Systems,
Springer, New York, 2006,

\bibitem{Angelo} J. A. Angelo and D. Buden, Space Nuclear Power, Orbit Book
Company, Malabar, FL, (1985).

\bibitem{8} https://www.lockheedmartin.com/en-us/products/compact-fusion.html

\bibitem{17} S. Cohen, et al. Direct fusion drive for interstellar
exploration, JBIS \textbf{72}, 37-50 (2019).

\bibitem{9} J. Cassibry, et al., Case and development path for fusion
propulsion, J. Spacecraft and Rockets \textbf{52}, 595 (2015).

\bibitem{10} S. A. Cohen, et al., Reducing neutron emission from small
fusion rocket engines, In: 66th International Astronautical Congress, IAC
2015, Jerusalem, Israel, Paper \# IAC-15,C4.7-C3.5,9,x28852.

\bibitem{101} M. Augelli, G.F. Bignami, G. Genta, Fission Fragments Direct
Heating for Space Propulsion -- Programme Synthesis and Applications to
Space Exploration, Acta Astronautica, \textbf{82}, 153-158 (2013).

\bibitem{11} T. J. Dolan, Magnetic Fusion Technology, Lecture Notes in
Energy, T. J. Dolan (ed.), Springer-Verlag London, 2013.

\bibitem{12} Y. S. Razin, G. Pajer, M. Breton, E. Ham, J. Mueller, M.
Paluszek, A. H. Glasser, S. A. Cohen, A direct fusion drive for rocket
propulsion, Acta Astronautica \textbf{105}, 145--155 (2014).

\bibitem{13} V. I. Khvesyuk, N. V. Shabrov, A. N. Lyakhov, ASH pumping from
mirror and toroidal magnetic confinement systems, Fusion Tech. 27, 406--408
(1995).

\bibitem{14} M. Sawan, S. Zinkle, J. Sheffield, Impact of tritium removal
and 3He recycling on structure damage parameters in a D-D fusion system,
Fusion Eng. Des. 61-62, 561--567 (2002).

\bibitem{15} W. Bang, et all. Temperature measurements of fusion plasmas
produced by petawatt laser-irradiated D2-3He or CD4-3He clustering gases,
Phys. Rev. Lett. \textbf{111}, 055002 (2013).

\bibitem{Jones} I. R. Jones, A review of rotating magnetic field current
drive and the operation of the rotamak as a field-reversed configuration and
a spherical tokamak, Phys. Plasmas \textbf{6}, 1950 (1999).

\bibitem{Cohen2007} S. A. Cohen, B. Berlinger, C. Brunkhorst, et al.,
Formation of collisionless high-$\beta $ plasmas by odd-parity rotating
magnetic fields, Phys. Rev. Lett. \textbf{98}, 145002 (2007).

\bibitem{Petrov} Y. Petrov, X. Yang, Y.Wang, and T-S Huang, Experiments on
rotamak plasma equilibrium and shape control, Phys. Plasmas \textbf{17},
0112506 (2010).

\bibitem{Guo} H. Y. Guo, M.W. Binderbauer, T. Tajima, et al., Achieving a
longlived high-beta plasma state by energetic beam injection, Nature
Communications, \textbf{6,} 6897 (2015).

\bibitem{Cohen2000} S. A. Cohen and R. D. Milroy, Maintaining the closed
magnetic-fieldline topology of a field-reversed configuration with the
addition of static transverse magnetic fields, Phys. Plasmas \textbf{7},
2539 (2000).

\bibitem{16} M. Paluszek, G. Pajer, Y. Razin, J. Slonaker, S. Cohen, R.
Feder, K. Griffin, M. Walsh, Direct Fusion Drive for a Human Mars Orbital
Mission, In: 65th International Astronautical Congress, IAC 2015, Toronto,
Canada, Paper \# IAC-14,C4,6.2.

\bibitem{Fick} A. Fick, Phil. Mag., \textbf{10}, 30 (1855).

\bibitem{ThomasOhio} S. J. Thomas, M. A. Paluszek, S. A. Cohen, A. Glasser,
Nuclear and future flight propulsion - modeling the thrust of the Direct
Fusion Drive, AIAA Propulsion and Energy Forum, 2018 Joint Propulsion
Conference, Cincinnati, Ohio, July 9-11, 2018.

\bibitem{UEDGE} T. Rognlien, J. L. Milovich, M. E. Rensink, G. D. Porter,
The UEDGE Code, J. Nucl. Mat., \textbf{196--198}, 347 (1992).

\bibitem{18} D. A. Shea, D. Morgan The Helium-3 shortage: supply, demand,
and options for congress. CRS Report for Congress. Congressional Research
Service 7-5700, R41419, www.crs.gov

\bibitem{Wittenberg} L. J. Wittenberg, J. F. Santarius, G. L. Kulcinski,
Lunar Supply of $^{3}$He for commercial fusion power, J. Fusion Techn.
\textbf{10}, 167 (1986).

\bibitem{19} E. N. Slyuta, A. M. Abdrakhimov, E. M. Galimov, V. I.
Vernadsky, The estimation of Helium-3 probable reserves in lunar regolith.
Lunar and Planetary Science \textbf{38}, 2175 (2007).

\bibitem{20} F. H. Cocks, $^{3}$He in permanently shadowed lunar polar
surfaces. Icarus. \textbf{206}, 778--779 (2010).

\bibitem{Wittenberg2} 15 L.J. Wittenberg, E.N. Cameron, G.L. Kulcinski, et
al., A Review of Helium-3 resources and acquisition for use as fusion fuel,
J. Fusion Techn. 21, 2230 (1992).

\bibitem{Robert} R. G. Kennedy, The interstellar fusion fuel resource base
of our solar system, JBIS, \textbf{71}, 298--305 (2018).

\bibitem{21} C.H. Williams, L.A. Dudzinsky et al., Realizing
\textquotedblleft 2001: A Space Odyssey\textquotedblright : Piloted
Spherical Torus Nuclear Fusion Propulsion, 37th Joint propulsion Conference
and Exhibit, Salt lake City, 2001, NASA/TM-2005-213559.

\bibitem{Thomas2019} S. J. Thomas, M. Paluszek, S. A. Cohen. Fusion-enabled
pluto orbiter and lander. Technical report, Princeton Satellite Systems,
Inc., 2019.

\bibitem{Genta} G. Genta, R.Ya. Kezerashvili, Achieving the required mobility
in the solar system through Direct Fusion Drive, 8th CSA-IAA Conference on
Advanced Space Technology, Shanghai, China, September 2019.

\bibitem{22} G. Genta, P. F. Maffione, Comparison Between Different
Approaches to Interplanetary Mission Design, International Journal of Signal
Processing, \textbf{2}, 2017, p. 54-66.

\bibitem{23} F. Chang D\'{\i}az, J. Carr et al., Solar Electric Propulsion
for Human Mars Missions, Acta Astronautica, Vol. 160, July 2019, p. 183-194

\bibitem{24} G. Genta, P. F. Maffione, Interplanetary Missions Performed
Outside the Optimal Launch Windows, 68th International Astronautical
Congress, Adelaide, Sept. 2017.
\end{thebibliography}
\end{document}